\documentclass[english,aps,reprint,superscriptaddress]{revtex4-1}
\usepackage[latin9]{inputenc}
\usepackage{color}
\usepackage{textcomp}
\usepackage{amsmath}
\usepackage{graphicx}

\makeatletter
 
 \@ifundefined{textcolor}{}
 {%
   \definecolor{BLACK}{gray}{0}
   \definecolor{WHITE}{gray}{1}
   \definecolor{RED}{rgb}{1,0,0}
   \definecolor{GREEN}{rgb}{0,1,0}
   \definecolor{BLUE}{rgb}{0,0,1}
   \definecolor{CYAN}{cmyk}{1,0,0,0}
   \definecolor{MAGENTA}{cmyk}{0,1,0,0}
   \definecolor{YELLOW}{cmyk}{0,0,1,0}
 }

\makeatother

\usepackage{babel}
\begin{document}
\begin{abstract}
Currently the half-life of $^{195}$Os is listed as unknown in most databases because the value of the only available measurement had been reassigned. We argue that the original assignment is correct and re-evaluate the half-life of $^{195}$Os to be 6.5(11)~min, consistent with the original measurement.  We also suggest to reassign the half-life of $^{195}$Ir to 2.29(17)~h.
\end{abstract}

\title{Reexamining the half-lives of $^{195}$Os and $^{195}$Ir}

\author{M. Birch}
\email{birchmd@mcmaster.ca}
\affiliation{Department of Physics and Astronomy, McMaster University, Hamilton, Canada}

\author{J. Flegenheimer}
\affiliation{flegen@arnet.com.ar, Buenos Aires, Argentina}

\author{Z. Schaedig}
\affiliation{National Superconducting Cyclotron Laboratory, Michigan State University, East Lansing, MI 48824, USA}

\author{B. Singh}
\affiliation{Department of Physics and Astronomy, McMaster University, Hamilton, Canada}

\author{M. Thoennessen}
\affiliation{National Superconducting Cyclotron Laboratory, Michigan State University, East Lansing, MI 48824, USA}
\affiliation{Department of Physics and Astronomy, Michigan State University, East Lansing, MI 48824, USA}

\maketitle

Basic properties of neutron-rich nuclei along the N~=~126 isotones are important for the astrophysical r-process (see for example \cite{Cow04}). However, below the doubly magic stable nucleus $^{208}$Pb they are very difficult to produce. While $^{207}$Tl \cite{Hah08} and $^{206}$Hg \cite{Nur61} have been known for a long time and the first half-life measurement of $^{205}$Au was reported in 1994 \cite{Wen94}, even lighter isotones became accessible only recently. The discoveries of $^{204}$Pt, $^{203}$Ir, $^{202}$Os as well as a few additional isotopes beyond N~=~126 were made possible by the development of improved separation techniques at the FRS fragment separator at GSI \cite{Ste08,Alv10,Mor11,Kur12}.

The half-life of one specific N = 119 nucleus, $^{195}$Os, which one would expect to be known, is still controversial. While the Table of Isotopes lists a half-life of 6.5~min \cite{ToRI}, the majority of nuclear data bases and evaluations \cite{Zho99,Arb04,Rob12,Aud12,ENSDF,NUDAT,wallet} do not accept this value and quote only an approximate theoretical value of $\sim$9~min from gross theory of beta decay \cite{Tak73}. In addition, the half-life of the daughter nucleus $^{195}$Ir is also not well established. The current ENSDF data evaluation \cite{ENSDF} recommends a value of 2.5(2)~h which corresponds to an unweigthed mean of two measurements which do not agree with each other within in the quoted uncertainties \cite{Hof67,Jan68}.

Rey and Baro first deduced a half-life of 6.5~min for $^{195}$Os from the reaction $^{198}$Pt(n,$\alpha$) and identified the isotope from the decay of the known daughter nucleus $^{195}$Ir \cite{Rey57,Bar57,Rey58}. Although recently two high-spin isomeric states, a short-lived state of 34~ns \cite{Val04,Caa05,Ste11} and a long-lived state of $>$9~min \cite{Ree12} have been observed, there are no other measurements of the half-life of the $^{195}$Os ground state.

The non-acceptance of the half-life measurement by Rey and Baro is based on the apparent reassignment of the $^{195}$Ir daughter nucleus in a 1974 unpublished annual laboratory report by Colle {\it et al.}: ``\emph{Unfortunately, the then-existing assignment for $^{195}$Ir has subsequently been identified as $^{81}$Rb, arising from reactions induced in target impurities. As a result, the present assignment of $^{195}$Os will not withstand careful scrutiny}'' \cite{Col74}. The timeline in this argument by Colle {\it et al.} is incorrect. At the time of the Rey and Baro discovery of $^{195}$Os the accepted half-life for $^{195}$Ir was 140~min \cite{Chr52,But54}. A half-life of 2.3~h was also reported in 1961 \cite{homma} from measurement of beta and gamma activity. Only one year later, in 1962, was this value replaced by Claflin {\it et al.} who determined a half-life of 4.2~h from the ($\alpha$,p) reaction on a supposedly highly enriched $^{192}$Os target \cite{Cla62}. This was the measurement that was subsequently questioned by Hoffstetter and Daly who demonstrated that the enriched osmium target could have been contaminated by other elements and the observed half-life of 4.2~h actually resulted from either  $^{79}$Br($\alpha$,2n) or $^{81}$Br($\alpha$,4n) reactions and thus corresponded to $^{81}$Rb \cite{Hof67}. In addition, the measurement by Rey and Baro could not have suffered from the same contamination problem as the experiment by Claflin {\it et al.} because they did not use $\alpha$-induced reactions on enriched osmium targets but (n,$\alpha$) reactions on high-purity, natural platinum.

Thus we believe that Rey and Baro indeed observed the decay of $^{195}$Os. In order to extract the half-life of $^{195}$Os Rey and Baro included not only the growth and decay of the daughter $^{195}$Ir but also contributions from $^{193}$Os. Since the presently adopted half-lives for these isotopes differ from the values that Rey and Baro used in their fit \cite{ENSDF}, we refitted their data as presented in Figure 2 of Reference \cite{Rey57}. The fit contained three components: the decay of $^{195}$Os, the growth and decay of $^{195}$Ir, and the decay of $^{193}$Os.

For the half-life of $^{193}$Os the most recent value of 29.830(18)~h by Krane \cite{Kra12} was used. It should be mentioned that this value differs from the currently accepted value of 30.11(1)~h \cite{ENSDF,Ant92}. 

As mentioned earlier, the currently adopted half-life of $^{195}$Ir was deduced as the unweigthed average of two independent measurements: a 2.8(1)~h half-life reported by Hofstetter and Daly in 1968 \cite{Hof67} and a 2.3(2)~half-life measured by Jansen, Pauw, and Toeset a few months later \cite{Jan68}. The first value was obtained from an analysis of the 99~keV $\gamma$-ray from the decay of the first excited state in $^{195}$Pt daughter, assigning this $\gamma$-ray only to the ground state activity of $^{195}$Ir. However, Jansen {\it et al.} demonstrated that this state is also populated by the decay of the 3.8(2)~h isomeric state in $^{195}$Ir \cite{Jan68,Jan73}. Thus the value quoted by Hoffstetter and Daly is likely too high and should be discarded.
Jansen {\it et al.} took the contributions from both states into account and arrived at the value of 2.3(2)~h. This value was consistent with the first measurements of 140~min in the 1950's \cite{Chr52,But54} which were known to Rey and Baro at the time of their measurement of $^{195}$Os.

Present evaluations \cite{Zho99,Aud12,ENSDF,NUDAT,wallet} do not consider that Rey and Baro also independently measured the half-life of $^{195}$Ir and it presented in the same papers reporting the discovery of $^{195}$Os \cite{Rey57,Rey58}. They deduced a half-life of 2.2~h by chemically extracting iridium fractions following the decay of its parent $^{195}$Os. This decay most probably populated only the 3/2$^+$ ground state of $^{195}$Ir, rather than the 11/2$^-$ isomer, since the ground state  spin and parity of $^{195}$Os is expected to be 3/2$^-$ \cite{Ree12}. An isomer of half-life $>$9~min at 454 keV discovered in $^{195}$Os \cite{Ree12} with suggested spin-parity of 13/2$^+$ is not expected to be populated in the $^{198}$Pt(n, $\alpha$) reaction used by Ray and Baro \cite{Rey57}.

We digitized the data of Figure 2 of Ref. \cite{Rey58} displaying the decay curve of $^{195}$Ir and deduced a value of 2.29(17)~h from a least-squares fit. A similar analysis of Figure 3 of Ref. \cite{Rey58} gives a more precise half-life of 2.17(7)~h, however, because of possible contamination from other Ir isotopes in this decay curve, we prefer the data from Figure 2 of Ref. \cite{Rey58}. Hence, we recommend the value of 2.29(17)~h. We believe this represents the best and most reliable half-life of the $^{195}$Ir ground state and we have used this value in the fit of $^{195}$Os. This value agrees well with the result 2.3(2)~h from \cite{Jan68}, not with 2.8(1)~h from \cite{Hof67}.
 
\begin{figure}
\includegraphics[width=1\columnwidth]{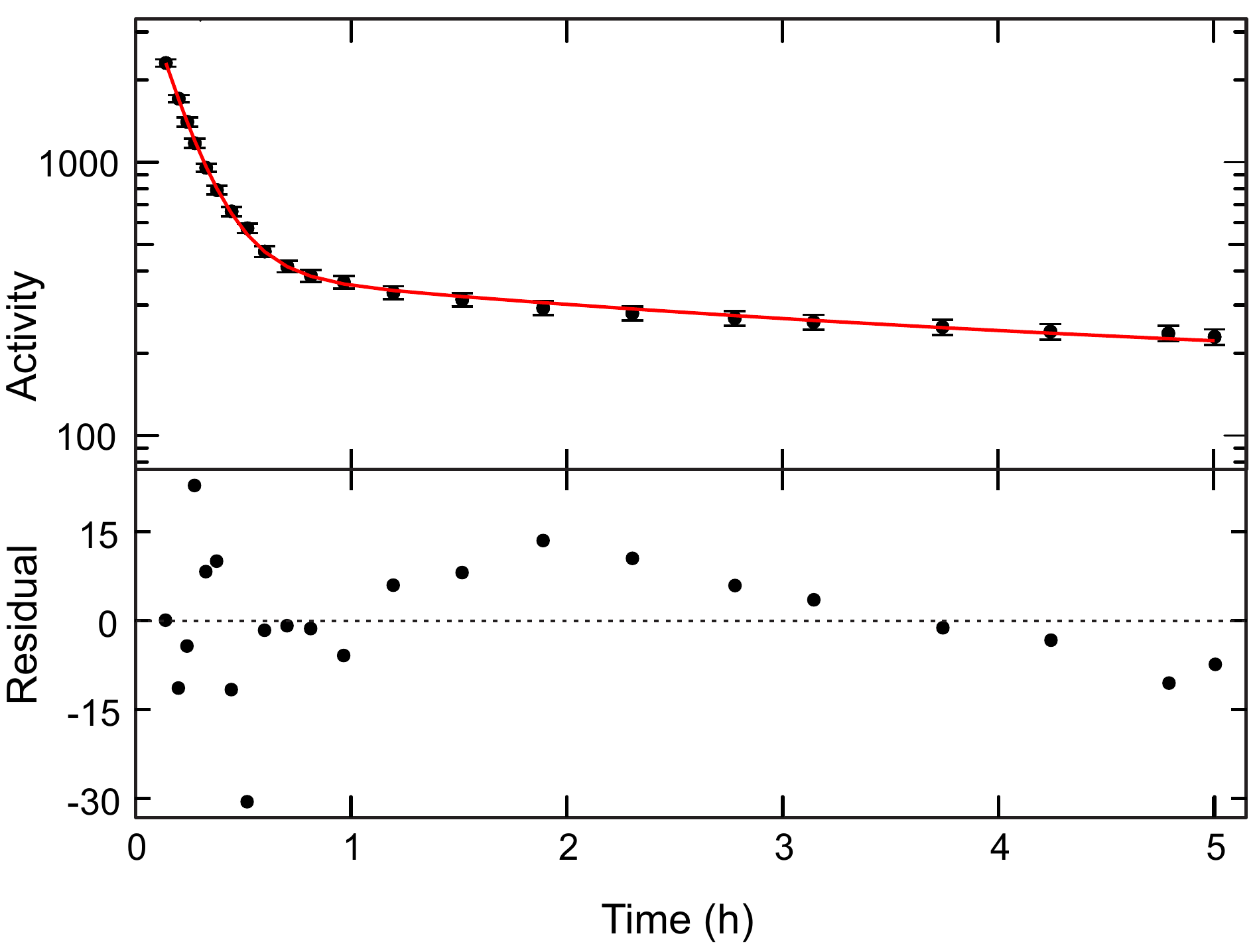}
\caption{(Color online) Decay curve and fit-residues for the decay of $^{195}$Os. The top panel shows our fit (solid red line) to the data of Figure 2 in Rey and Baro's work \cite{Rey57} (solid black circles) 
and the residuals of our fit are shown in the bottom panel.}
\label{fig:decay}
\end{figure}

Therefore, there remain four free parameters for the fit of the $^{195}$Os decay curve: the half-life of $^{195}$Os, and the initial amounts of $^{195}$Os, $^{195}$Ir, and $^{193}$Os. These four parameters were fitted by a least-squares method, where the
minimum sum of squared residuals was determined by differential evolution. The uncertainties in the fitted parameters were estimated by a Monte Carlo method in which many fits were performed on data sets generated from sampling
within the uncertainties of the data. Because the original paper did not give uncertainties we assigned
the statistical uncertainty given by $\sqrt{N}$ along with an uncertainty associated with
the digitization of the plot. The sample standard deviations of the set of fitted results from the
simulated data sets were taken to be the uncertainties in the best fit parameters. The results from
this procedure are shown in Figure \ref{fig:decay}. The deduced half-life for $^{195}$Os is 6.5(11)~min, in agreement with 6.5~min value quoted in the original Rey and Baro papers. In addition, we conclude that the half-life of the $^{195}$Ir ground state, based upon Rey and Baro's work, be accepted as 2.29(17)~h in contrast to 2.5(2)~h quoted in the evaluated databases \cite{ENSDF}. Furthermore, a new measurement of the $^{195}$Os ground state half-life using state-of-the-art techniques is highly desirable.

We would like to thank W. B. Walters for sending us a copy of the article ``Decay of $^{195}$Os and Search for $^{196}$Os'' in the 1974 University of Maryland Cyclotron Laboratory Progress Report \cite{Col74} and E. Browne for reading and evaluating the original article ``Un Nuevo Isotopo Del Osmio'' by Rey and Baro \cite{Rey57} and also for useful comments on our manuscript. This work was in part supported by the National Science Foundation under grant No. PHY11-02511 and by Office of Science, Office of Nuclear Physics of the U.S. Department of Energy.

\bibliographystyle{apsrev4-1}

\end{document}